# Kerr effect at high electric field in isotropic phase of mesogenic materials

Bing-Xiang Li,[1,2] Volodymyr Borshch,[1] Sergij V. Shiyanovskii,[1,*] Shao-Bin Liu,[2]

and Oleg D. Lavrentovich[1,†]

[1]*Liquid Crystal Institute and Chemical Physics Interdisciplinary Program,
Kent State University, Kent, Ohio, 44242, USA*
[2]*College of Electronic and Information Engineering, Nanjing University of
Aeronautics and Astronautics, Nanjing 210016, China*

**Abstract.** The well-known Kerr effect in isotropic fluids consists in the appearance of uniaxial orientational order and birefringence that grows as the square of the applied electric field. We predict and observe that at a high electric field, the Kerr effect displays new features caused by nonlinear dependence of dielectric permittivity on the field-induced orientational order parameter. Namely, the field-induced birefringence grows faster than the square of the electric field; the dynamics of birefringence growth slows down as the field increases. As a function of temperature, the field induced birefringence is inversely proportional to the departure from an asymptotic critical temperature, but this temperature is no longer a constant (corresponding to the lower limit of the supercooled isotropic phase) and increases proportionally to the square of the electric field.

Condensed matter in presence of external fields is usually described by a Landau type of the free energy expansion. The field-induced changes in the properties are assumed to be linearly related to each other. A good example is the field-induced orientational order in an isotropic fluid [1-8]. The phenomenon manifests itself through the field-induced birefringence (called Kerr effect for the electric field and Cotton-Mouton effect for the magnetic field) [1-7], or as an enhancement of dielectric permittivity [8]. The standard Landau-de Gennes model describes these effects by expanding the free energy density in power series of the orientational order parameter $S$, assuming that the field-induced properties such as birefringence and dielectric permittivity, are linearly proportional to $S$, namely, $\delta n \propto S$ and $\varepsilon_E = \varepsilon_{iso} + \varepsilon_1 S$, where $\varepsilon_{iso}$ and $\varepsilon_1$ are constants.

In this Letter, we demonstrate that for high applied fields, the response of the system can be properly described only when the field-induced parameters are related in a nonlinear fashion.

___________________________
* Corresponding author: sshiyano@kent.edu
† Corresponding author: olavrent@kent.edu

Namely, we show that the field induced dielectric permittivity $\varepsilon_E(S)$ exhibits a quadratic dependence on the field-induced scalar order parameter $S$, namely, $\varepsilon_E = \varepsilon_{iso} + \varepsilon_1 S + \varepsilon_2 S^2$, where the new coefficient $\varepsilon_2$ can be even larger than $\varepsilon_1$. The quadratic term leads to qualitatively new effects, such as birefringence growing faster than the square of the electric field, a slowing down of the response dynamics at high fields and dependence of the effective critical temperature on the applied electric field. All these predictions are confirmed experimentally.

**Model.** The equilibrium state of the isotropic phase in an electric field $\mathbf{E}$ is determined by the free energy density in the Landau-de Gennes model as:

$$f = \frac{1}{2}a(T-T^*)S^2 - \frac{1}{3}bS^3 + \frac{1}{4}cS^4 - \frac{1}{2}\mathbf{E}\cdot\mathbf{D}, \tag{1}$$

where $a$, $b$, and $c$ are the expansion constants, $T^*$ is the lower temperature limit of the supercooled isotropic phase, and the electric displacement $\mathbf{D} = \varepsilon_0 \varepsilon_E(S)\mathbf{E}$ depends on the applied field $\mathbf{E}$ directly and through the dependence of the dielectric permittivity $\varepsilon_E$ on $S$. The effect of external fields is usually considered weak and thus only the linear term in the expansion of $\varepsilon_E(S)$ is retained. As a result, the simplest version of the theory with $b=c=0$ predicts that the field-induced birefringence depends on temperature as $\delta n \propto (T-T^*)^{-1}$ [9,10]. The abundant experimental results, see e.g. [3,4,11-13], clearly validate this prediction, except for the close proximity of the isotropic-to-nematic transition $T_{NI}$ [14-16], where it suffices to use the full form of the Landau-de Gennes expansion with the higher order $b$- and $c$- terms in Eq.(1) [9,16]. The main result of our work is that the response of the system to high electric field is different from the predictions of the standard Landau-de Gennes model. This response demonstrates nonlinear

(quadratic) dependence of the dielectric permittivity on the orientational order, $\varepsilon_E = \varepsilon_{iso} + \varepsilon_1 S + \varepsilon_2 S^2$; the latter leads to qualitative new effects.

The equilibrium state corresponding to the minimum of $f$ obeys the condition

$$\frac{\partial f}{\partial S} = a(T-T^*)S - bS^2 + cS^3 - \frac{1}{2}\varepsilon_0(\varepsilon_1 + 2\varepsilon_2 S)E^2 = 0. \tag{2}$$

At temperatures well above $T_{NI}$, the field-induced order parameter $S$ is small, so the $b$- and $c$- terms in Eq.(2) can be neglected:

$$\frac{E^2}{S} = \frac{2a}{\varepsilon_0 \varepsilon_1}(T-T^*) - \frac{2\varepsilon_2}{\varepsilon_1}E^2. \tag{3}$$

The nonzero coefficient $\varepsilon_2$ in this expression shifts the asymptotic critical temperature for $S$,

$T' = T^* + \frac{\varepsilon_0 \varepsilon_2}{a} E^2$.

The dynamics of field-induced order parameter can be described in the Landau-Khalatnikov model $\gamma(dS/dt) = -\partial f / \partial S$ [17], obtained from Eq. (2) with $b = c = 0$ as

$$\gamma \frac{dS}{dt} = -\left[a(T-T^*) - \varepsilon_0 \varepsilon_2 E^2\right]S + \frac{1}{2}\varepsilon_0 \varepsilon_1 E^2. \tag{4}$$

For the square pulse, $E(t_{on} \leq t \leq t_{off}) = E$, the solution of Eq.(4) is

$$S(t_{on} \leq t \leq t_{off}) = S_E\left(1 - \exp\frac{t_{on}-t}{\tau_{on}}\right), \quad \tau_{on} = \frac{\gamma}{a(T-T^*) - \varepsilon_0 \varepsilon_2 E^2};$$

$$S(t > t_{off}) = S(t_{off})\exp\frac{t_{off}-t}{\tau_{off}}, \quad \tau_{off} = \frac{\gamma}{a(T-T^*)}, \tag{5}$$

where $S_E$ is the equilibrium value of the order parameter in the given applied electric field, obtained from Eq.(3); the switch-on time $\tau_{on}$ is a function of the applied electric field, while the switch off time $\tau_{off}$ is not. Interestingly and counterintuitively, the switch on time increases as the driving electric field increases. Below we demonstrate that the main predictions of this simple model, namely, the field dependence of the asymptotic critical temperature and field-triggered slow-down of the switch on response, are clearly observed in the experiment.

**Experiment.** We used the standard nematic 4-cyano-4'-pentylbiphenyl, purchased from Merck and Jiangsu Hecheng. The temperature of the isotropic to nematic transition phase transition is $T_{NI} = (35.4 \pm 0.1)\,°C$. The NLC is filled into a cell made of two parallel glass plates with thin transparent electrodes of indium tin oxide of small area, $2 \times 2\,\text{mm}^2$ and low resistivity, $10\,\Omega/\text{square}$. The cell thickness is $d = 6.5\,\mu\text{m}$. The glass plates were covered by layers of polyimide PI-1211 (Nissan). In order to measure the optic response to the applied field, we use a laser beam (He-Ne, $\lambda = 633\,\text{nm}$) that passes through the crossed polarizers with the cell and an optic compensator between them, as described previously [18]. The transmitted intensity is measured by a detector TIA-525 (Terahertz Technologies, response time $<1\,\text{ns}$). The cell is sandwiched between two right-angle prisms, so that the light incidence is oblique, at the angle $\theta = 45°$, Fig.1 (a). The temperature of cell assemblies is controlled with Linkam LTS350 hot stage with the accuracy better than $0.1\,°C$. Voltage pulses with sharp rise and fall edges (characteristic time better than 3 ns) were applied by a pulse generator HV 1000 (Direct Energy). The voltage pulses and photodetector signals were monitored with the 1G samples/s digital oscilloscope TDS2014 (Tektronix), Fig. 1(b).

The field induced birefringence was measured by monitoring the dynamics of polarized light intensity transmitted by the cell, optical compensator and the pair of polarizers. The dynamics of the transmitted light intensity is measured at two compensator's settings, A and B, for which the phase retardance difference is $\Gamma_B - \Gamma_A = \pi$. When there is no electric field, the measured transmitted light intensities for these settings are $I_A(0) = I_B(0) = \frac{I_{max} + I_{min}}{2}$; here $I_{max}$ and $I_{min}$ are the maximum and minimum transmitted intensities determined by adjusting the phase retardance $\Gamma$ of the compensator. Then the effective field-induced birefringence $\delta n_{eff}(t)$ is determined through field-induced phase retardance $\Delta\Gamma$ as (see [18] for details):

$$\delta n_{eff}(t) = \frac{\lambda \Delta \Gamma}{2\pi d} = \frac{\lambda}{2\pi d} \arcsin \frac{I_A(t) - I_B(t)}{I_{max} - I_{min}} \qquad (6)$$

Using the Fresnel equation for the experimental set-up, we determine that

$$\delta n_{eff} = n_\perp \sqrt{1 - n_g^2 \sin^2\theta / n_\parallel^2} - \sqrt{n_\perp^2 - n_g^2 \sin^2\theta} = A\left(n_\parallel^2 - n_\perp^2\right) = ABS, \qquad (7)$$

where $A = n_g^2 \sin^2\theta \Big/ \left(n_\parallel n_\perp \sqrt{n_\parallel^2 - n_g^2 \sin^2\theta} + n_\parallel^2 \sqrt{n_\perp^2 - n_g^2 \sin^2\theta}\right)$, $n_\parallel$ and $n_\perp$ are field modified refractive indices that correspond to polarizations parallel and perpendicular to the applied field, respectively, $n_g = 1.52$ is the refractive index of the glass prism, and $\theta = 45°$ is the incidence angle, Fig.1 (a). With the experimental data $n_\parallel = 1.58$, $n_\perp = 1.55$, collected at $E = 8.8 \times 10^7$ V/m and $T = T_{NI} + 25°C$, we find $A = 0.21$; the latter number remains constant within 3% when the applied field is less than $1.2 \times 10^8$ V/m. According to the Vuks model for the local field correction

[19] and to the experimental data [20,21], the quantity $n_\parallel^2 - n_\perp^2$ in Eq.(7) is proportional to the field induced order parameter $S$; $n_\parallel^2 - n_\perp^2 = BS$.

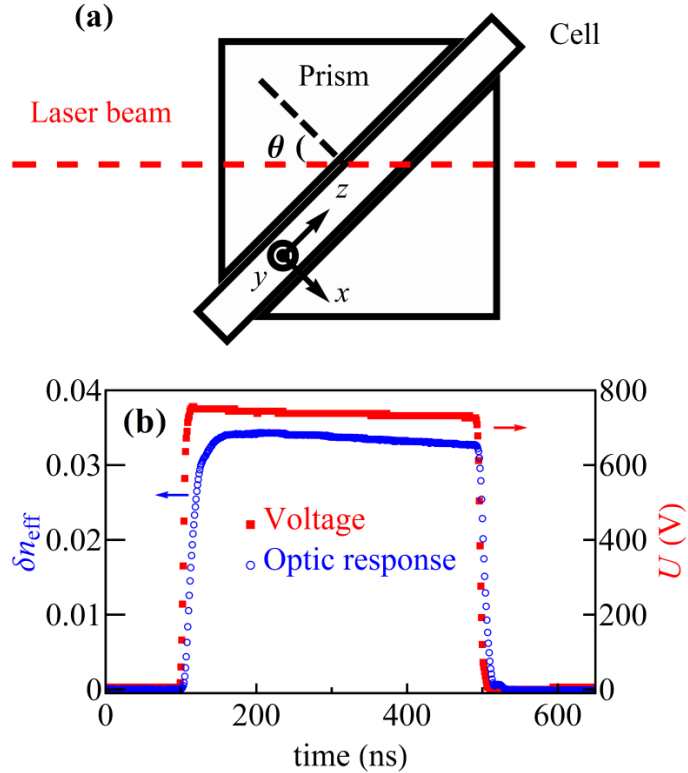

**FIG. 1 (color online). (a) Experimental incidence geometry, in which the cell is sandwiched between two right angle prisms for oblique light incidence at 45 degrees; the electric field is applied along the x axis; (b) dynamics of effective field-induced birefringence (circles) response to the applied voltage pulse (squares) at 30ºC above $T_{NI}$.**

The measured dependencies $\delta n_{eff}(E,T)$ allow us to verify Eq.(3), see Fig.2,3. Far above the transition temperature, for $T > T_{NI} + 25°C$, the temperature dependence of $E^2/\delta n_{eff}$ is clearly linear for all values of the strong electric field, Fig.2a, as predicted by Eq.(3). The intersection of the linear dependencies with the horizontal temperature axis in Fig.2a is different for different voltages. This intersection is the asymptotic critical temperature $T'$ introduced above. The explicit field dependence of $T'$, obtained by fitting the data in Fig.2a for $T > T_{NI} + 25°C$, follows the behavior $T' = T^* + \frac{\varepsilon_0 \varepsilon_2}{a} E^2$, Fig.2b, predicted by the model. Small deviations from the linear behavior $T' - T^* \propto E^2$ observed at the highest fields will be discussed later (see Fig.3). Note that the very strong electric fields used in our work (that were not available in the prior studies of the subject) cause such a significant increase of $S$ that the temperature dependencies of $E^2 / \delta n_{eff}$ close to the transition, $T_{NI} < T < T_{NI} + 25°C$, is no longer linear, Fig.2a, and thus must be described with the higher order ($b$ and $c$) terms in the Landau-de Gennes expansion.

Equation (3) predicts that $E^2 / S$ is proportional to $E^2$ at the temperatures well above $T_{NI}$; this is indeed what is observed experimentally, Fig. 3(a). The proportionality constant in the relationship $n_\parallel^2 - n_\perp^2 = BS$ is estimated to be $B = 1.0 \pm 0.1$ from the temperature dependencies $n_\parallel(T)$, $n_\perp(T)$ and $S(T)$ measured in the nematic phase. One can estimate that the field-induced order parameter in the fitting temperature range can reach 0.15; however, this value is still mainly determined by linear and dielectric terms in Landau-De Gennes model, Eq.(2), as the relative contribution of the non-linear $b$ and $c$ terms is less than 4 %. Fitting the data at high temperatures allows one to determine the temperature behavior of $a(T - T^*)/\varepsilon_1$ and $\varepsilon_2/\varepsilon_1$, Fig.3b. The slope of $a(T - T^*)/\varepsilon_1$ vs $T$ in Fig. 3(b) leads to $a/\varepsilon_1 = 2.0 \times 10^4 \text{J/°C m}^3$. The ratio $\varepsilon_2/\varepsilon_1$ increases from 0.43 to 1.04 when the temperature decreases from $T_{NI} + 51°C$ to $T_{NI} + 25°C$. The temperature dependence of $\varepsilon_2/\varepsilon_1$ is most probably caused by the coefficient $\varepsilon_2$, associated with intermolecular interactions, rather than by $\varepsilon_1$ describing single molecule additive contributions. Note that the finite slope of the $E^2/S$ vs $E^2$ means that the field-induced birefringence $\delta n \propto S$

*does not* follow the classic dependence $\delta n \propto E^2$ of the Kerr effect; in our case, $\delta n$ grows faster than $E^2$.

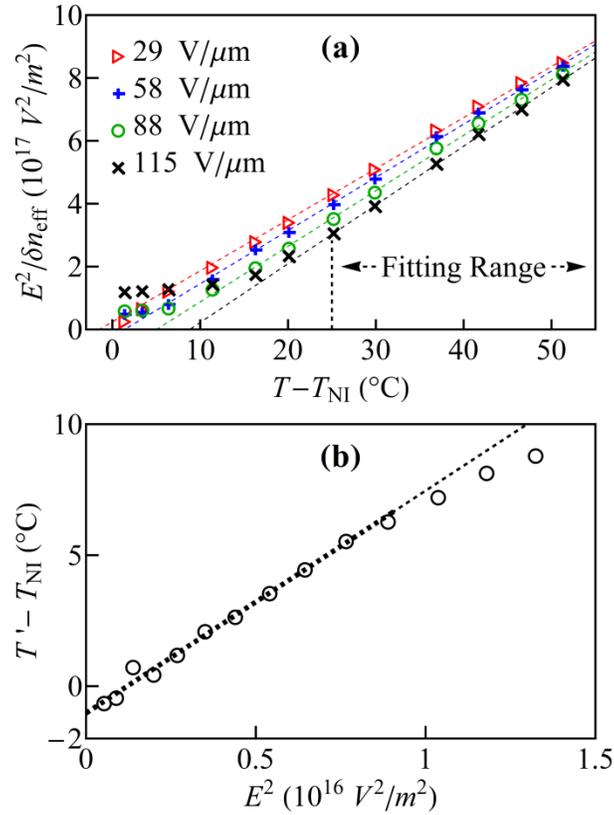

**FIG. 2 (color online). (a) Temperature dependence of** $E^2/\delta n_{eff}$ **for applied electric field** $E = 29$ V/$\mu$m **(triangles),** $58$ V/$\mu$m **(pluses),** $88$ V/$\mu$m **(circles), and** $115$ V/$\mu$m **(saltires); (b)** $T' - T_{NI}$ **vs.** $E^2$. **All dashed lines show the corresponding results of the linear fitting performed for high temperatures,** $T > T_{NI} + 25°C$.

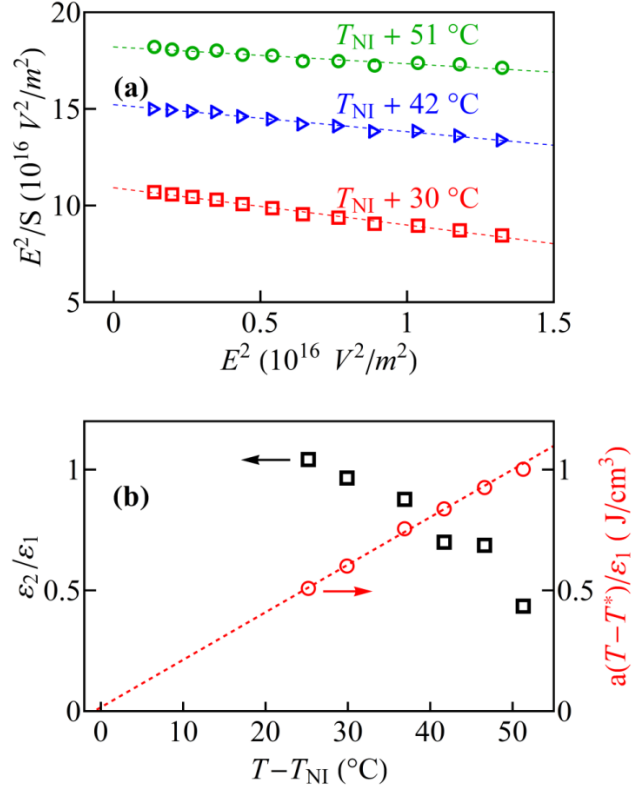

**FIG. 3 (color online). (a)** $E^2/S$ **vs.** $E^2$ **at** $T_{NI}+30°C$ **(squares),** $T_{NI}+42°C$ **(triangles),** $T_{NI}+51°C$ **(circles); (b) Temperature dependence of** $\varepsilon_2/\varepsilon_1$ **(squares) and** $a(T-T^*)/\varepsilon_1$ **(circles).**

The significant contribution of the quadratic term in $\varepsilon_E(S)$ can be qualitatively explained from the temperature dependence of dielectric and optic tensors in the nematic phase. The applied electric field creates a uniaxial paranematic phase with the optical axis parallel to the field. Thus, we compare $\varepsilon_E(S)$ with the parallel component of the dielectric permittivity in the nematic phase $\varepsilon_\parallel(S) = \varepsilon_{iso} + \varepsilon_1^\parallel S + \varepsilon_2^\parallel S^2$. The latter dependence can be reconstructed from the temperature dependencies $\varepsilon_\parallel(T)$ and $S(T)$ deduced from the dielectric and birefringence measurements. Our measurements of $\varepsilon_\parallel(T)$ and $\delta n(T)$, similar to ones in [22,23], result in $\varepsilon_1^\parallel = 6.4$ and $\varepsilon_2^\parallel = 8.8$.

The ratio $\varepsilon_2^\| / \varepsilon_1^\| \approx 1.38$ obtained for the nematic phase, is close to $\varepsilon_2 / \varepsilon_1$ measured for the Kerr effect in the isotropic phase.

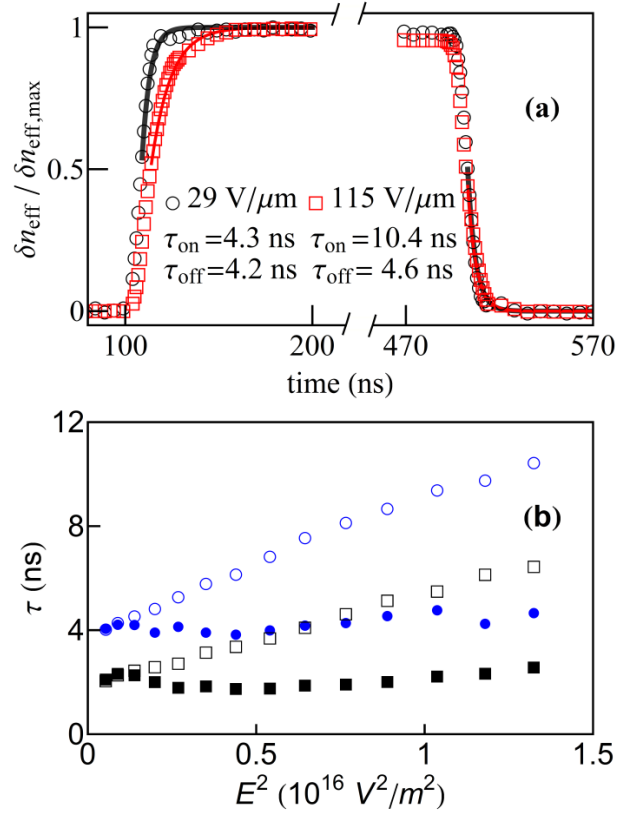

**FIG. 4 (color online). Dynamics of the switching in isotropic phase at the temperatures well above $T_{NI}$: (a) experimental optic response at $T_{NI} + 30°C$ (circles) and the respective linear fitting (solid line) for two different applied electric fields; note the slower switching-on at higher fields; (b) Electric field dependence of $\tau_{on}$ (open symbols) and $\tau_{off}$ (closed symbols) at $T_{NI} + 30°C$ (circles) and $T_{NI} + 42°C$ (squares).**

We now proceed to the discussion of the dynamics of field-induced response. It takes a finite time (about 10 ns) for the voltage to change from zero to its saturated value; the same for the reverse process. Therefore, the values of $\tau_{on}$ and $\tau_{off}$ were obtained by fitting the rise and fall processes within the time intervals that correspond to the saturated and zero voltages, respectively. To compare the dynamics at different electric field, in Fig.4a we plot the normalized values of the field induced birefringence, $\delta n_{eff} / \delta n_{eff,max}$, where $\delta n_{eff,max}$ is the maximum birefringence achieved at the given field. Figure 4(a) clearly shows that $\tau_{on}$ increases as the field is increased; $\tau_{off}$ does not depend on the field within the accuracy of the experiment (1 ns), being approximately equal to $\tau_{on}$ at the small field, Fig. 4(b), as expected, see Eq.(5).

The experimental results above are all explained within the proposed model of the Kerr effect at high electric fields. The model advances the standard Landau-de Gennes theory by adding a dielectric permittivity term proportional to the square of the induced order parameter. An important question is whether the data can be explained by other mechanisms. One possibility is to add a term $\propto SE^4$ with a fourth order electric field to the Landau-de Gennes expansion. Introduction of such a term would produce dependencies $E^2 / \delta n_{eff}$ vs. $T - T_{NI}$ that have a different tilt but the same intersection with the temperature axis for different values of $E$; such a behavior would contradict strongly the experimental data in Fig.2(a). Furthermore, the term $\propto SE^4$ would not make the switch-on time dependent on the applied electric field. One can also consider heating effects caused by adiabatic changes of polarization and order parameter [24] and by Joule heating of liquid crystal material and electrodes at the substrates. The resulting temperature increase is rather small, less than 0.2°C, and could only decrease the observed asymptotic critical temperature and make the switching-on time faster. The experiments, however, show the opposite behavior, Figs.2 and 4.

To conclude, we presented a theoretical description and experimental confirmation of the new features of the electrooptic Kerr effect, observed at high electric fields. First, at a given temperature, the field-induced birefringence grows faster than $E^2$, which is of interest in both the fundamental and applied aspects. Second, the rise-on time of the field-induced birefringence becomes longer as the field increases. Finally, at the fixed electric field, the temperature dependence of the inverse Kerr constant outside the close proximity of $T_{NI}$ remains linear, but the asymptotic critical temperature $T'$ does not coincide with $T^*$, shifting upwards with the square of the electric field. When the temperature dependence of the inverse Kerr constant is used to determine the lower temperature limit of the isotropic phase, this effect should be accounted for, otherwise the procedure will yield wrong results. All these features are new as compared to the standard Landau-de Gennes description of the Kerr effect and underline the importance of nonlinear relationships between different field-induced properties.

**Acknowledgements.** The work was supported by NSF grants DMR-1507637 and DMR-1410378, State of Ohio through Ohio Development Services Agency and Ohio Third Frontier grant TECG20140133. B.-X. Li acknowledges China Scholarship Council support.